\documentclass[article]{aa} 
\usepackage{graphicx}
\usepackage{epsfig}
\begin{document}

\title{Proper motions and velocity asymmetries in the RW Aur jet\thanks{Based
on observations collected at the Canada-France-Hawaii 
 Telescope (CFHT), operated by the National Research Council of Canada, the
 Centre National de la Recherche Scientifique, and University of Hawaii.}
 }

\author{Luis L\'opez-Mart\'{\i}n\inst{1,2}\and 
Sylvie Cabrit\inst{2}\and Catherine Dougados\inst{3}}

\institute{Instituto de Astrof\'\i sica de Canarias,
c/ V\'\i a L\'actea s/n 38200, La Laguna, Tenerife, Spain
({\tt luislm@ll.iac.es})
\and LERMA, Observatoire de Paris, UMR 8112 du CNRS, 
61 Av. de l'Observatoire, F-75014 Paris \and Laboratoire d'Astrophysique,
Observatoire de Grenoble, UMR 5571, BP 53, F-38041 Grenoble Cedex 9}

   \date{Received; accepted}
   
   \abstract{We present adaptive optics spectro-imaging observations of the
   RW~Aur jet in optical forbidden lines, at an angular resolution of
   0.4$''$. Comparison with HST data taken 2 years later shows that proper
   motions in the blueshifted and redshifted lobes are in the same ratio as
   their radial velocities, a direct proof that the velocity asymmetry in
   this jet is real and not an emissivity effect. The inferred jet
   inclination to the line of sight is $i = 46\pm 3$\degr. The inner knot
   spacing appears best explained by time variability with at least two
   modes: one irregular and asymmetric (possibly random) on timescales of
   $\le $3-10 yr, and another more regular with $\simeq$ 20 yr period. We
   also report indirect evidence for correlated velocity and excitation
   gradients in the redshifted lobe, possibly related to the blue/red
   velocity and brightness asymmetry in this system.
}

\authorrunning{L\'pez-Mart\'{\i}n et al.}
\titlerunning{Proper motions in the RW Aur jet}
\maketitle

\section{Introduction}

\noindent

Jets from young stars are believed to be powered by accretion and
accelerated and collimated by magnetic forces (see e.g. K\"onigl \& Pudritz
\cite{ppiv} for a review). However, a clear identification of the magnetic
launch configuration is still needed to fully understand the role of jets
in the physics of accreting protoplanetary disks. The optically revealed
innermost jet regions of T~Tauri stars (TTS) hold essential clues to this
question, and are the subject of growing observational efforts at
sub-arcsecond resolution (e.g. Lavalley-Fouquet et al. \cite{lavalley};
Dougados et al. \cite{paper1} (hereafter Paper I); Bacciotti, Mundt, Ray et
al. \cite{bacc}).
HH~229, the bipolar jet from the actively accreting star RW~Aur~A, is a
particularly interesting target. It was first identified in long-slit
spectroscopy by Hirth et al. (\cite{hirth}), who noted an asymmetry in
radial velocity between the blueshifted and redshifted lobes of a factor
$\simeq$1.8. Further study with HST/STIS (Woitas et al. \cite{woitas},
hereafter Paper~II) traces this asymmetry down to 0.1$''$ $\approx$ 14 AU
from the source, suggesting that it might be an intrinsic property of the
jet driving engine. In addition, the RW~Aur jet contains many emission
knots, traced as far as 50$''$-100$''$ (Mundt \& Eisl\"offel \cite{mundt})
down to only 0.1$''$-10$''$ from the star (Paper I, Paper~II), which may be
used to test models of knot formation in stellar jets.

In this Letter, we present a ground-based spectro-imaging study of the
RW~Aur jet that brings several new results, including improved knot proper
motions, jet inclination, and knot dynamical times, as well as indirect
evidence for internal velocity and excitation gradients. Our observations
are summarized in Sect.~\ref{Observations} and our results are discussed in
Sect.~\ref{Results}.

\section{Observations}\label{Observations}

Observations were carried out at the CFHT on December 18-20, 1998 with the
integral field spectrograph OASIS. We used a spatial sampling of
0.16$''$/lens and achieved an angular resolution of 0.4$''$ FWHM after
adaptative optics correction.  Two spectral ranges were observed (6213-6534
\AA\ and 6496-6824 \AA) with a resolving power $\approx $ 90 km s$^{-1}$. A
second dataset with 0.11$''$/lens was obtained over a smaller field of
view.
Following standard OASIS reduction (cf. Lavalley-Fouquet et
al. \cite{lavalley}), a specific procedure was applied to subtract at each
position the strong continuum+emission spectrum of RW~Aur~A, scaled by the
PSF, and to construct datacubes of residual forbidden line profiles. More
details are given in Dougados et al. (2003, in preparation). Comparison
between the two lens samplings shows that continuum-subtracted emission
maps at 0.16$''$/lens are reliable beyond 0.35$''$ from the star, while the
0.11$''$/lens data are more reliable at $d < 0.35''$.  Radial velocities
are computed in the stellar reference frame (heliocentric velocity +16 km
s$^{-1}$; Petrov, Gahm, Gameiro  et al. \cite{petrov}). We adopt hereafter the notation
[OI]$\equiv$[OI]$\lambda$6300, [SII]$\equiv$[SII]$\lambda$6731, and
[NII]$\equiv$[NII]$\lambda$6583.

\section{Results}\label{Results}
\noindent

\subsection{Proper motions of redshifted knots}\label{proper-red}
\noindent

\begin{figure}
\resizebox{\hsize}{!}{\includegraphics{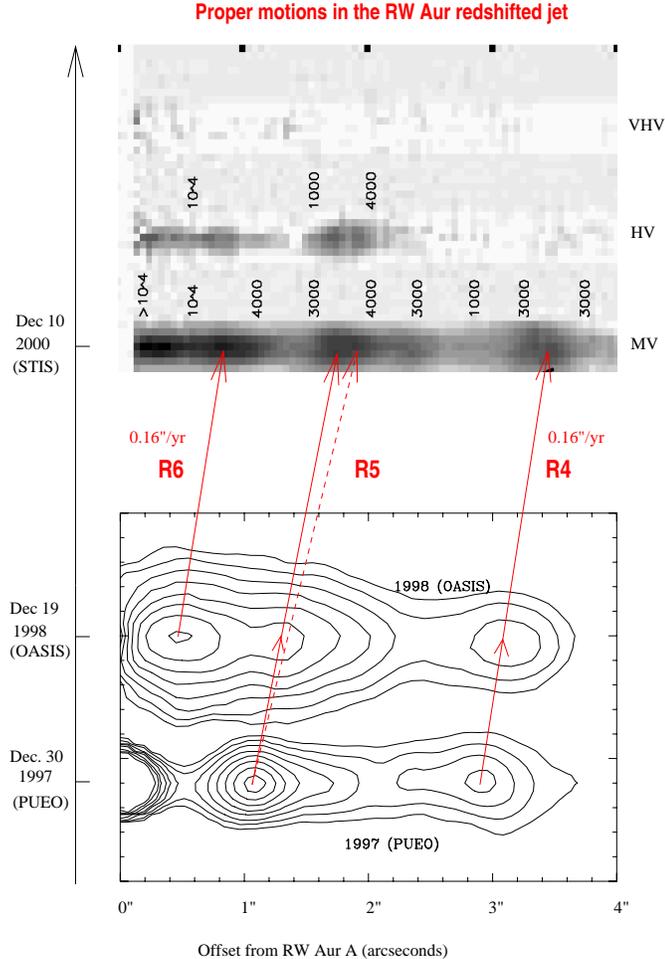}}
\caption{Illustration of knot proper motions in the inner redshifted jet of
RW~Aur.  From top to bottom: [S II] channel maps obtained in Dec. 2000 with
STIS/HST (Paper~II); integrated [S II] image obtained in
Dec. 1998 with OASIS at 0.16$''$/lens (this paper); deconvolved [O I] image
obtained in Dec. 1997 with PUEO at CFHT (Paper~I).}
\label{properred}
\end{figure}

Figure \ref{properred} compares our 1998 [S~II] image of the
redshifted jet with images obtained in Dec. 2000 (Paper~II)
and in Dec. 1997 (Paper~I)\footnote{In order to replicate
the correct separation between RW~Aur~A and B, the Dec. 1997 image
published in Paper~I had to be corrected for a small
5\% underestimate in the FOCAM pixel size}. A smooth systematic shift
is seen between the 3 epochs, indicative of proper motions. Table
\ref{table1} lists the knot positions and the resulting proper motions
in arcsec per year. As the OASIS spatial resolution is not sufficient to
detect possible offsets between different lines, we adopt for the nominal
1998 position of each knot in Table~1 an average of independent estimates
using the strongest two lines ([OI] and [SII]) and the two lens samplings
(0.11$''$ and 0.16$''$). Knots are numbered from the outside in, starting
at the most distant one detected by Dougados et al.~(Paper I).
For the outermost knot seen with OASIS (knot R4), the 3 epochs are
consistent with a constant proper motion of 0.16$''$ yr$^{-1}$. The proper
motion for the following knot (R5) is more uncertain, as it appears double
in the 2000 STIS image. The range of possible values is 0.17$''$-0.27$''$
yr$^{-1}$. For knot R6, we find again 0.16$''$ yr$^{-1}$, based on the 1998 and
2000 epochs only. We note that an innermost emission peak is detected in
the 0.11$''$/lens dataset at $\simeq$ 0.28$''$. It may correspond
to the peak at 0.25$''$ in the STIS image, and would then be stationary.

\begin{table*}
\begin{flushleft}
\center
\begin{tabular}{|c|c|c|c|c|c|c|c|}
\hline
Knot &$d_{97}$ & $d_{98}$ & $d_{00}$ & $|V_{\rm rad}|$ & $\mu$ & $i$ & $t_{98}-t_{ej}$ \\
& ($''$) & ($''$) & ($''$) & (km s$^{-1}$) &($''$ yr$^{-1}$) & (\degr) & (yr) \\
\hline
\hline
\multicolumn{8}{|c|}{\bf Red lobe}\\ 
\hline
   &      &$\le$0.28 & 0.25  &100-{\it 96}  & stationary? &  & \\
R6    &        & 0.51   & 0.82      &94-{\it 117}  & 0.16 & 44$\pm$8  & 1.3 \\
R5   &   1.07:& 1.32   & 1.72/1.91 &107-{\it 133} &0.24$\pm$0.05&53$\pm$9& 8/5\\
R4   &   2.92 & 3.12   & 3.45      &97-{\it 100}  & 0.16  & 47$\pm$5 & 18 \\
R3   & 3.9-4.6&        &           &       & [0.16]     &       & 24-29 \\
R2    & 7.9    &        &           &       & [0.16]     &       & 49 \\
R1   & 11.1   &        &           &       & [0.16]     &       & 69 \\
\hline
\multicolumn{8}{|c|}{\bf Blue lobe}\\ 
\hline
   & & $\le$0.19 & 0.2 & 175-{\it 168}& stationary? & & \\
B4  & & & 0.57 & {\it 179}& [0.26] & & 0.2 \\
B3 & & 0.73$\pm 0.05$ & 1.25 & 178-{\it 168}& 0.26$\pm 0.035$ & 45$\pm$5 & 3 \\
B2   & 2.8 &         &       &       & [0.26]&    & 12 \\
B1    & 7.1 &         &       &       & [0.26]&    & 28 \\
\hline
\end{tabular}
\end{flushleft}
\caption[] {Properties of knots in the RW Aur jet. $d_{97}$, $d_{98}$,
and $d_{00}$ are the projected knot offsets in Dec. 1997, 1998, and 2000
(Paper~I; this paper; Paper~II), with respective uncertainties $\pm
0.01''$, $\pm 0.02''$, and $\pm 0.05''$, unless when quoted. $V_{\rm rad}$ is
the knot [S~II] centroid velocity from this paper (roman font) and from
Paper~II (italics); $\mu$ the proper motion (typical uncertainty $\pm
0.027''$yr$^{-1}$; values between brackets are assumed); $i$ the angle to the
line of sight (error bars include uncertainties in both $\mu$ and
$V_{\rm rad}$); $t_{98}-t_{ej}$ the time between knot ``ejection'' and Dec. 20,
1998.}

\label{table1}
\end{table*}

\subsection{Proper motions of blueshifted knots}\label{proper-blue}

\begin{figure}
\resizebox{\hsize}{!}{\includegraphics[height=8cm]{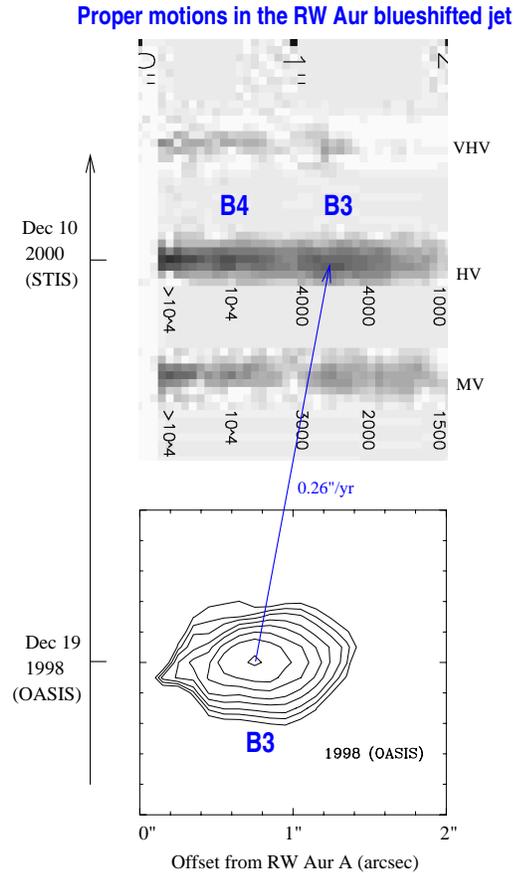}}
\caption{Comparison of [S~II] images of the RW~Aur blueshifted jet obtained
with OASIS in Dec.~1998 (this paper) and with STIS/HST in Dec.~2000
(Paper~II), illustrating the proper motion.}
\label{properblue}
\end{figure}

Figure \ref{properblue} compares [S~II] images of the inner blueshifted jet
in 1998 (this paper) and 2000 (Paper~II).  We lack a 1997 epoch,
as deconvolution residuals in the PUEO image dominate over the faint blue
jet emission out to $\simeq 2''$ (Paper I). We find a proper
motion of 0.26$''$ yr$^{-1}$ for the main knot (B3). A new knot (B4) has
appeared at 0.6$''$ in the 2000 STIS image. With a similar proper motion
to B3, it would have been within 0.05$''$ of the star at the date of the
OASIS observations and thus unresolvable. Finally, our 0.11$''$/lens
dataset shows an innermost peak at $\simeq 0.19''$, which
may correspond to the 0.2$''$ peak seen in the STIS image and would then be
stationary.

\subsection{Revised jet inclination}\label{inclination}

Assuming that knot proper motions trace actual fluid motions and not wave
pattern speeds, the angle of inclination $i$ of the jet with respect to the
line of sight is given by $\tan(i) = (\mu/'' {\rm yr}^{-1})\times
(666.7/V_{\rm rad})$, where $V_{\rm rad}$ is the velocity projected on the line of
sight, and a distance of 140 pc to Taurus-Auriga is assumed.  We adopt as
$V_{\rm rad}$ for each knot the centroid of a gaussian fit to the [S~II] line
profile. Values from this paper and from Paper~II (in italics) are
given in Col. 5 of Table \ref{table1}. Discrepancies in $V_{\rm rad}$ between
the two datasets are discussed in more detail in Section
\ref{velocity}. The resulting values of $i$ are given in Col. 7, with an
uncertainty interval that includes both the uncertainty in $V_{\rm rad}$ and
that in $\mu$ (see notes to Table~1). Inclination values are remarkably
consistent between knots, within the estimated errors. In particular, we
derive consistent inclinations for the blue and red lobes, i.e. the proper
motion in the blue jet is higher than in the red jet by the same factor
($\simeq 1.6$) as their average centroid radial velocities.  This fact
argues strongly for a real velocity difference between the blueshifted and
redshifted jets, rather than an apparent difference due solely to slow
bowshock wings dominating the emission around the redshifted jet (proper
motions should then be similar in the two lobes despite differing radial
velocities). A weighted average over both lobes yields\footnote{The larger
inclination value $i = 53$\degr\ found by Woitas et al. (Paper~II) for knot R4
stems from a proper motion estimate 20\% too high, inferred from the
published PUEO image of Paper~I before correction for pixel
scale (see footnote 1)} $i= 46$\degr$\pm$3\degr.

\subsection{Knot spacing, dynamical times, and origin}\label{ejection}
\noindent
The RW~Aur observations allow to measure knot spacings and dynamical
timescales very close to the source and thus bring interesting
constraints on the origin of knots in stellar jets. To increase the
distance and time span in our analysis, we include in Table~1 more
distant knots imaged with PUEO in 1997, which lie beyond the OASIS
field of view and were not considered by Woitas et al. (Paper~II).

Table~\ref{table1} shows that the knot spacing first increases with
distance from the star, from $0.7''$-$0.8''$ at $d\simeq 1.5''$ (knots
R5-R6, B3-B4) to $1.8''$-$2.4''$ at $d\simeq 3''$ (knots R4-R5, B2-B3), and
then levels off at $\Delta x \simeq 3''-4''$ for $d\geq 4''$.  Adopting as
the jet characteristic radius $r_{\rm j}$ half the FWHM measured at the same
distances in Paper~II, the inner knot spacing is then $\Delta x\simeq
8-10 \, r_{\rm j}$. For comparison, the axisymmetric pinch instabilities in MHD jets
simulated by Lery et al. (2000) produce jet knot spacings of $\Delta x/r_{\rm j}
\leq 3$, while Kelvin-Helmholtz instabilities predict $\Delta x/r_{\rm j} \simeq
V_{\rm j}/c_{\rm s} > 10$.  An even more important observational constraint is that the
first moving knot appears at a distance $d_{\rm min} < \Delta x$, while the
growth scale of an instability is expected to be several times its
wavelength. Hence, neither kind of instability seems adequate to explain
the innermost knot spacings, although a range of MHD jet configurations
should be explored before they are fully ruled out.

Another possible origin for the knots, which can easily yield $d_{\rm min}
< \Delta x$, is time variability in the jet ejection speed (cf. Equ. (3) in
Raga et al. \cite{raga1}). The last column of Table \ref{table1} gives the
dynamical ages of each knot assuming constant proper motion (with $t=0$
arbitrarily set at our OASIS observation date of Dec.19,~1998).  For knots
lacking proper motion measurements, we adopt the typical $\mu$ in the same
lobe (in brackets in Col.~6 of Table~\ref{table1}). We find that the
timescale between knots initially increases with distance, from 3 to 10
yr, and then appears to level off at $\simeq$ 20 yr beyond
4$''$. Furthermore, the inner knots do not appear to be ``synchronized''
between the two jet sides, while there is tentative evidence of
synchronization when the 20 yr period appears (knots R3 and B1). We
conclude that there would be two distinct processes of time variability in
RW~Aur: one that is irregular {\it and} asymmetric --- possibly random
(random time variability of the jet speed produces asymptotically an
increase in knot spacing with distance; Raga \cite{raga2}) and another more
regular and symmetrical with a 20 yr period --- possibly related to a
stellar magnetic cycle. We suggest that asynchronous variability might
occur, e.g., through local magnetic reconnexion events that affect only one
jet side.

The apparently stationary knots at $0.2''$-$0.3''$ = 30-45 AU from the star
need to have a different origin, possibly recollimation shocks (e.g. Gomez de
Castro \& Pudritz \cite{gomez}) or temperature gradients in a progressively
heated wind (e.g. Garcia et al. \cite{garcia}), but predicted distances are
presently too model-dependent to yield useful constraints.

\subsection{Velocity gradients and velocity asymmetry}\label{velocity}

\begin{figure}
\resizebox{\hsize}{!}{\includegraphics[angle=0]{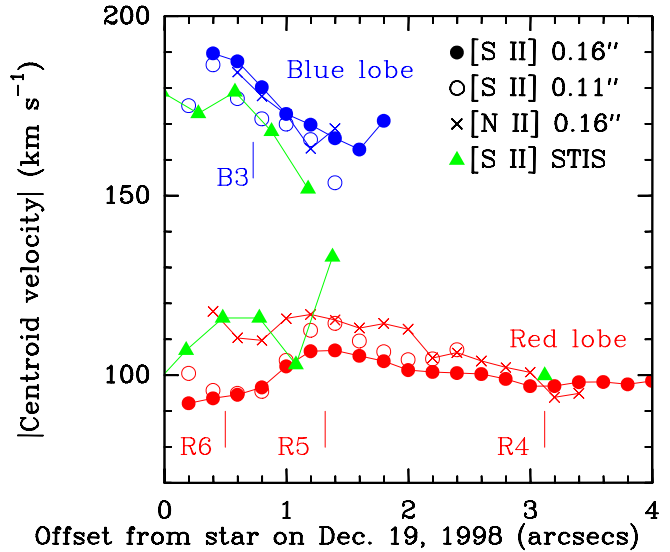}}
\caption{Centroid radial velocity in [S II]
(from gaussian fits) as a function of distance from the star. Filled
  and open circles refer to the 0.16$''$/pixel and 0.11$''$/pixel OASIS
  data, respectively, integrated over a 1$''$ wide slit. For comparison,
  crosses show centroids in [N II]
  (0.16$''$/lens), and triangles show [S II] centroids in a
  0.1'' slit with HST/STIS (Woitas et al. 2002), shifted to correct for
  proper motion (by -0.32$''$ in the red lobe, -0.52$''$ in the blue;
  cf. Table~1).}
\label{figuraanchos}
\end{figure}

Figure \ref{figuraanchos} plots the [S II] centroid radial velocity as a
function of distance from the star for the OASIS and STIS datasets (the
latter is shifted to correct for proper motions). In the blue jet, the
various datasets agree to within 10 km $\rm s^{-1}$, comparable to expected
instrumental and calibration errors. The situation is different in the
redshifted jet: the 0.11$''$/lens and 0.16$''$/lens datasets still agree
with each other within $\le$ 10 km $\rm s^{-1}$, but they lie typically 20 km $\rm s^{-1}$ below
the STIS values out to the position of knot R5.  Further out, however, at
the position of knot~R4, all data converge to a very similar $V_{\rm rad}$.

These discrepancies can only be understood if the faster gas seen on the
redshifted jet axis by STIS (0.1$''$ wide slit) suffers from beam-smearing
in the OASIS data. In other words, sharp transverse velocity gradients must
occur on scales $\leq 0.4''$ within the redshifted jet, out to distances $<
3''$ from the star\footnote{such gradients seem confirmed by
preliminary analysis of transverse STIS slits (Woitas et al., work in
progress)}. We note that [N~II] centroids from OASIS (shown as crosses in
Fig.~\ref{figuraanchos}) agree better with the [S~II] STIS centroids,
despite an angular resolution identical to the [S~II] OASIS data. This fact
further indicates that the faster axial gas contributes a greater fraction
of the total flux in [N~II] than in [S~II], i.e. that it has higher
excitation.  Interestingly, we find no evidence for such velocity and
excitation gradients in the blueshifted jet. If confirmed, this
difference might be related to the blue/red velocity asymmetry between the
two jet lobes; e.g., bowshock wings or turbulent boundary-layers could be
somehow enhanced in the slower lobe (thus also explaining its higher
brightness). Further theoretical and observational work is needed to
explore this possibility.

\acknowledgements LLM is grateful to Observatoire de Paris and to
Minist\`ere de l'Education Nationale for a postdoctoral fellowship.

\clearpage

\end{document}